\documentclass[showpacs,amsmath,amssymb,aps,10pt,reprint,superscriptaddress,prb]{revtex4-1}
\usepackage{graphicx}
\usepackage{bm}
\usepackage[breaklinks=true,colorlinks=true,linkcolor=blue,urlcolor=blue,citecolor=blue]{hyperref}
\usepackage{graphics}
\usepackage{dcolumn}
\usepackage{amsmath,amssymb}
\usepackage{mathdots}
\usepackage{natbib}
\usepackage{soul,color}
\usepackage{epstopdf}
\usepackage[note-name]{notes2bib}

\begin{document}
\title{Magnon-Fluxon interaction in a ferromagnet/superconductor heterostructure}

\author{O.~V.~Dobrovolskiy}
    \affiliation{Physikalisches Institut, Goethe University, 60438 Frankfurt am Main, Germany}
    \affiliation{Physics Department, V. Karazin Kharkiv National University, 61077 Kharkiv, Ukraine}
\author{R. Sachser}
    \affiliation{Physikalisches Institut, Goethe University, 60438 Frankfurt am Main, Germany}
\author{T. Br\"acher}
\author{T. Fischer}
    \affiliation{Fachbereich Physik and LFZ OPTIMAS, Technische Universit\"at Kaiserslautern, Germany}
\author{V.~V.~Kruglyak}
    \affiliation{School of Physics and Astronomy, University of Exeter, EX4 4QL Exeter, UK}
\author{R. V. Vovk}
    \affiliation{Physics Department, V. Karazin Kharkiv National University, 61077 Kharkiv, Ukraine}
\author{V.~A.~Shklovskij}
    \affiliation{Physics Department, V. Karazin Kharkiv National University, 61077 Kharkiv, Ukraine}
\author{M. Huth}
    \affiliation{Physikalisches Institut, Goethe University, 60438 Frankfurt am Main, Germany}
\author{B. Hillebrands}
\author{A. V.~Chumak}
    \affiliation{Fachbereich Physik and LFZ OPTIMAS, Technische Universit\"at Kaiserslautern, Germany}

\begin{abstract}
Ferromagnetism and superconductivity are most fundamental phenomena in condensed matter physics. Entailing opposite spin orders, they share an important conceptual similarity: Disturbances in magnetic ordering in magnetic materials can propagate in the form of spin waves (magnons) while magnetic fields penetrate superconductors as a lattice of magnetic flux quanta (fluxons). Despite a rich choice of wave and quantum phenomena predicted, magnon-fluxon coupling has not been observed experimentally so far. Here, we clearly evidence the interaction of spin waves with a flux lattice in ferromagnet/superconductor Py/Nb bilayers. We demonstrate that, in this system, the magnon frequency spectrum exhibits a Bloch-like band structure which can be tuned by the biasing magnetic field. Furthermore, we observe Doppler shifts in the frequency spectra of spin waves scattered on a flux lattice moving under the action of a transport current in the superconductor.
\end{abstract}

\maketitle

\section{Introduction}
\enlargethispage{2\baselineskip}
Superconductivity is a physical phenomenon which has attracted large interest over many years from a fundamental as well as from an applied point of view\cite{Rog11boo}. In 1957, Alexei Abrikosov predicted the formation of vortices of supercurrent, referred to as Abrikosov vortices or fluxons, which can explain the peculiar response of a type-II superconductor to a magnetic field\cite{Shu37etf,Shu08ujp,Abr57etp}. Fluxons form a lattice in the superconductor, where each of them represents a whirl of supercurrent shielding the superconductor from the magnetic field emanating from and attaining a maximum at the vortex core\cite{Abr03nol,Bra95rpp,Mos10boo,Dob17pcs}. It seems natural to bring a type-II superconductor in contact with a ferromagnetic material which hosts spin waves, the Goldstonde modes of the spin systems\cite{Blo30zph,Gur96boo}. Magnons, the quanta of spin wave excitation, by themselves have attracted large interest, for instance, as a model system for Bose-Einstein condensation\cite{2016arXiv161207305S,Boz16nph,Dem06nat} or as beyond-CMOS data carriers in the research fields of magnonics and magnon spintronics\cite{Chu15nph,LENK2011107,Kru10jpd}. In addition, magnons are excellent probes, since they offer an unmatched sensitivity for the surroundings and the interfaces of a magnetic material\cite{Gru85pss,PhysRevB.41.530,Dik15apl}. In a ferromagnet-type-II-superconductor heterostructure, magnon-fluxon interaction has been predicted to give rise to many interesting phenomena \cite{Bar96prl,Ngx98prb,Bul05prl,Lin12prb,She11prl,Bes13prb,Bes14prb}, such as magnon radiation by Abrikosov vortices\cite{Bes14prb} and an associated enhancement of the vortex viscosity due to this radiation of spin waves\cite{She11prl}. Experimentally, however, magnon-fluxon interaction has not been observed so far. In this work, we demonstrate magnon-fluxon interaction experimentally. We show that it enables the convenient all-electrical detection of the presence and the periodicity of the Abrikosov lattice by its characteristic fingerprint in the magnon spectrum, since the lattice constitutes a reconfigurable magnonic crystal\cite{Kra14pcm,Chu17jpd}. We also demonstrate that the magnon spectrum is affected by the motion of the vortex lattice which causes a Doppler effect, immediately suggesting tunable spin wave devices and the electrical detection of the vortex motion with high precision.

\section{Results and Discussion}

\begin{figure}
    \centering
    \includegraphics[width=0.9\linewidth]{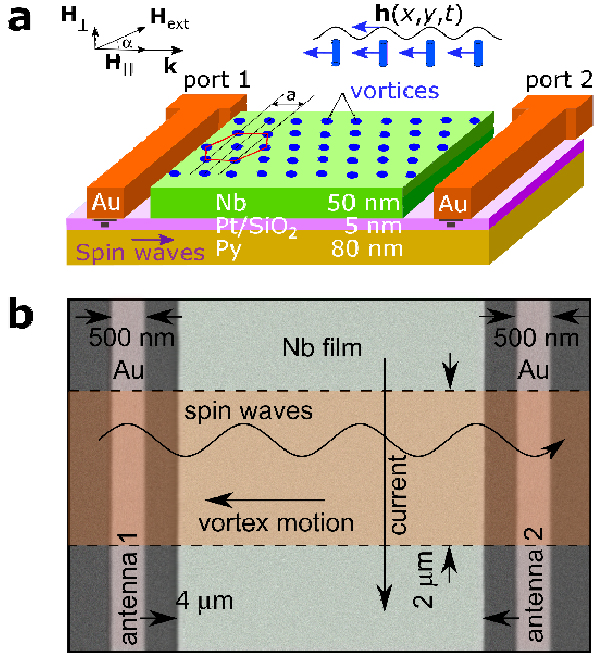}
    \caption{\textbf{Experimental system}.
    \textbf{a} Sketch and \textbf{b} false-color scanning electron microscope image of the Nb and Au layout.  The Py/Nb bilayer is in an inclined magnetic field $\mu_0 \mathbf{H}_{ext}$ with constant in-plane component $\mu_0 H_{||}= 59.5$\,mT and out-of-plane component $\mu_0 H_\perp$ varying between $3$ and $11$\,mT. Backward volume magnetostatic spin waves are excited by antenna 1, propagate through the Py waveguide, whose arrangement is indicated by the dashed lines in \textbf{b}, and are detected by antenna 2. The vortex lattice induces a spatially periodic magnetic field $\mathbf{h}(x,y)$ in Py, which becomes alternating in time when the vortices move.
    \label{f1}
}
\end{figure}

\textbf{Experimental system.} For our proof-of-concept experiments we investigate a bilayer system consisting of a $80\,\mathrm{nm}$-thick ferromagnetic (F) Permalloy (Py) in contact with a $50\,\mathrm{nm}$ thick superconducting Nb film (S), Fig. \ref{f1} a. At this thickness of the Nb, the presence of in-plane vortices\cite{Vla16prb} can be excluded. The Nb bridge and the microwave antennae were positioned perpendicular to the Py spin wave waveguide as discussed in the methods section and as is indicated in Fig. \ref{f1} b. The assembly was mounted in a cryogenic high-frequency holder to conduct measurements in the vortex state of Nb at 8\,K.

Each fluxon carries a quantum of magnetic flux $\Phi_0 = h/2e = 2.07\times10^{-15}$\,Tm$^2$ and they may roughly be viewed as small cylinders whose cores are in the normal state and that are arranged in a hexagonal lattice. Furthermore, the fluxons can be regarded as tiny whirls of supercurrent, producing local magnetic field variations of about $2\mu_0 H_{c1}$. Here, $\mu_0 H_{c1}$ is the lower critical field of the Nb film, for which we estimate $2$\,mT at $T = 8$\,K. These fields extend over distances on the order of $2\lambda$ in the plane perpendicular to the vortex axis, where $\lambda$ is the magnetic penetration depth\cite{Bra95rpp}. With the zero-temperature estimate $\lambda(0)\simeq100$\,nm\cite{Gub05prb} and using the two-fluid expression\cite{Kim03cry} $\lambda(T) = \lambda(0)[1 - (T/T_c)^4]^{-1/2}$, we obtain $\lambda(8\,\mathrm{K})\approx150$\,nm for our films.

The Py film acts as host for spin wave propagation. By applying a microwave current to the excitation antenna (port 1) of the structure, spin waves are excited in the Py waveguide by the antenna's oscillating Oersted field. After their propagation through the waveguide they are detected at port 2. The bilayer is placed in an inclined magnetic field $\mathbf{H}_\mathrm{ext}$ whose components serve different purposes. The in-plane component of the field is aligned along the Py waveguide and magnetizes it along its long axis. This field-component sets the spin wave propagation to the backward volume magnetostatic spin wave (BVMSW) configuration and defines the frequency gap of the spin wave spectrum\cite{Gur96boo}. The field is adjusted for this component to be equal to $\mu_0 H_{||} = 59.5\,\mathrm{mT}$ in all experiments. The vertical component $\mu_0 H_\perp$ is used to set-up a vortex lattice in the Nb as it is cooled below the critical temperature $T_c$ and undergoes the phase transition into the S state\cite{Shu37etf,Abr57etp,Shu08ujp}. It is varied from $3\,\mathrm{mT}$ to $11\,\mathrm{mT}$, resulting in different lattice spacings.

\begin{figure*}
    \centering
    \includegraphics[width=0.8\linewidth]{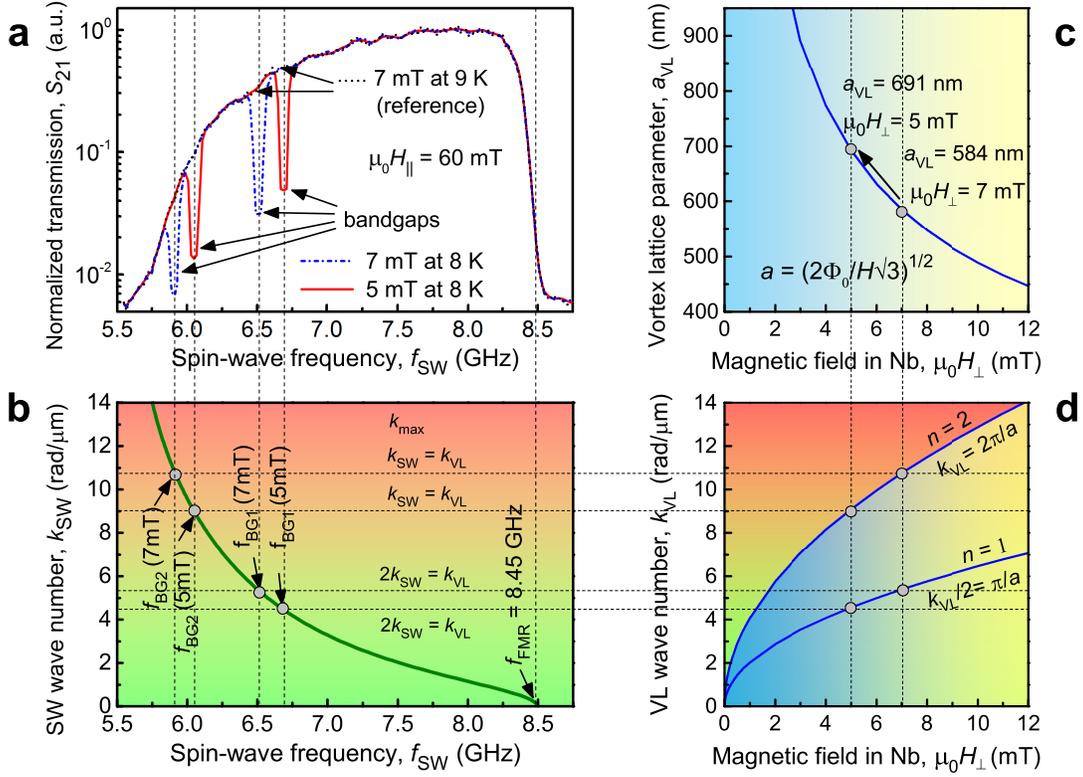}
    \caption{\textbf{Fluxon-induced reconfigurable magnonic crystal}.
    \textbf{a} Bandgaps in the spin wave transmission spectra for a series of out-of-plane magnetic field components $\mu_0 H_\perp$.
    \textbf{b} Magnetic field dependence of the vortex lattice parameter $a_\mathrm{VL}(H_\perp) = (2\Phi_0/\sqrt{3}H_\perp)^{1/2}$.
    \textbf{c} The spin wave dispersion relation $k_\mathrm{SW}(f_\mathrm{SW})$ calculated within the framework of the Kalinikos-Slavin theory adapted to a thin-film waveguide \cite{Kal86jpc,Bra17phr}.
    \textbf{d} Magnetic field dependence of the wave vector of the vortex lattice $k_\mathrm{VL}(H_\perp) = \pi (2\sqrt{3}H_\perp/\Phi_0)^{1/2}$ and $\frac{1}{2}k_\mathrm{VL}(H_\perp)$.
    The horizontal dashed lines indicate the Bragg resonance conditions $k_\mathrm{SW} = \frac{1}{2}nk_\mathrm{VL}(H_\perp) = \frac{1}{2} n\pi(2\sqrt{3} H_\perp/\Phi_0)^{1/2}$ linking the bandgap frequencies $f_{\mathrm{BG}n}(H_\perp)$ with the vortex lattice parameter $a_\mathrm{VL}(H_\perp)$.
    \label{f2}
}
\end{figure*}

\textbf{Fingerprint of the vortex state in the spin-wave spectrum.} From the viewpoint of the spin wave system, the local fields created by the vortex lattice constitute a Bragg grating with partial reflection at each vortex. Such a structure constitutes a magnonic crystal and features specific bandgaps in its transmission spectrum, which is analogous to phononic or photonic crystals\cite{Kra14pcm,Chu17jpd,Lu2009,doi:10.1080/09500349414550261}. This implies that the spin wave spectrum will feature characteristic dips in its transmission that are directly linked to the wave vector of the fluxon lattice. This is demonstrated in Fig. \ref{f2}~a, where the transmission spectrum $S_{21}$ from port 1 to port 2 is shown in the presence (red and blue curves) and in the absence (black curve) of the vortex lattice. The measurement in the absence of vortices acts as reference. It has been recorded with a vertical field of $\mu_0 H_\perp = 7\,\mathrm{mT}$ at $9\,\mathrm{K}$, i.e., just above the phase transition into the normal state where no vortices are present. It resembles a typical BVMSW transmission spectrum\cite{Chu10prb,Kra14pcm}. Figure \ref{f2} b displays the (inverse) dispersion relation $k_\mathrm{SW}(f_\mathrm{SW})$ for the BVMSW spin waves, which was calculated within the framework of the Kalinikos-Slavin theory adapted to a thin-film waveguide\cite{Kal86jpc,Bra17phr} for a magnetic field $\mu_0 H_{\parallel} = 59.5$\,mT and a saturation magnetization of $M_\mathrm{s} = 675\,$kA/m, which has been obtained from ferromagnetic resonance spectroscopy measurements. 
A gyromagnetic ratio of $\gamma = 28$\,GHz/T and an exchange stiffness constant of $A = 1.6\times10^{-11}$\,J/m have been assumed. In the dispersion relation, lower frequencies correspond to larger wave numbers due to the negative group velocity in the BVMSW geometry\cite{Kra14pcm,Chu10prb}. The dispersion relation explains the main features of the reference transmission spectrum in Fig. \ref{f2} a. The steep roll-off on the high-frequency side corresponds to the ferromagnetic resonance (FMR) frequency of $f_\mathrm{FMR} = 8.45$\,GHz. The gentler slope on the low-frequency side can be understood by taking into account the finite size of the microwave antenna. For the given antenna width of $500$\,nm, an efficient excitation of spin waves is restricted to spin waves up to a maximum wave vector $k = 2\pi/w = 12.6$\,rad/$\mu$m. According to Fig. \ref{f2} b, this results in a lower limit for the spin wave frequency slightly below $f_\mathrm{min} \approx 5.8$\,GHz. The excitation conditions for spin waves depending on their wave vector are summarized in the figure by the shading: Close to the FMR, i.e., in the green region at wave vectors close to zero, spin waves can be excited efficiently by the antenna, whereas towards the red area the excitation efficiency drops\cite{doi:10.1002/pssb.201147093}.

When the Nb film is in the normal state N, the Py waveguide behaves like a usual spin wave conduit. In contrast, the spin wave transmission exhibits characteristic changes if the Nb is cooled down into the S state. The blue curve in Fig. ~\ref{f2} a shows the corresponding transmission spectra at $8\,\mathrm{K}$ and $\mu_0 H_\perp = 7\,\mathrm{mT}$. Characteristic dips are visible in the spin wave transmission, situated at $f_\mathrm{BG1} \approx 6.5\,\mathrm{GHz}$ and $f_\mathrm{BG2} \approx 5.92\,\mathrm{GHz}$, each featuring a full width at half depth of approximately $\Delta f_\mathrm{BG}  = 50$\,MHz. These gaps are the direct consequence of the Bragg scattering at the vortex lattice. To illustrate this, Fig.~\ref{f2} c shows the theoretically expected vortex lattice parameter $a_\mathrm{VL}$ given by $a_\mathrm{VL} =  (2\Phi_0/\sqrt{3}H_\perp)^{1/2}$ for the assumed triangular vortex lattice\cite{Bra95rpp}. Here, the shading represents the quality of the lattice: In the region of low fields, i.e., blue shading, the lattice is not very uniform and the spacing is irregular due to defects. For larger fields towards the yellow area, the uniformity of the lattice improves. Specifically, $\mu_0 H_\perp = 7$\,mT results in $a_\mathrm{VL} = 584$\,nm. With the general Bragg condition $2a_\mathrm{VL} = n\lambda_\mathrm{SW}$, which corresponds to $2k_\mathrm{SW} = n k_\mathrm{VL}$, it can be seen that the frequency positions of the dips, which correspond to the wave vectors $k_\mathrm{SW1}=5.4$\,rad/$\mu$m and $k_\mathrm{SW2}=10.8$\,rad/$\mu$m, belong to the first and the second order Bragg scattering, respectively. To illustrate this further, the red curve in Fig. \ref{f2} a shows the transmission spectrum if the film is first heated up to the N state at $9\,\mathrm{K}$, the field is changed to $\mu_0 H_\perp = 5\,\mathrm{mT}$, and the sample is again cooled into the S state. Now, the lattice parameter changes to $a = 691\,\mathrm{nm}$. Consequently, the bandgaps are located at $f_\mathrm{BG1} \approx 6.7\,\mathrm{GHz}$ and $f_\mathrm{BG2} \approx 6.1\,\mathrm{GHz}$, corresponding to wave vectors of $\pi/a = 4.55\,\mathrm{rad}/\mu\mathrm{m}$ and $2\pi/a = 9.1\,\mathrm{rad}/\mu\mathrm{m}$, respectively.

The wave number of the vortex lattice is given by
\begin{equation}
\label{e1}
    k_\mathrm{VL} = \pi (2\sqrt{3}H_\perp/\Phi_0)^{1/2}.
\end{equation}
and, thus, the matching condition for the wave vectors of spin waves and the vortex lattice reads
\begin{equation}
\label{e2}
    k_\mathrm{SW}(f_{\mathrm{BG}n}) = n\frac{k_\mathrm{VL}(H_\perp)}{2} =  \frac{n\cdot\pi (2\sqrt{3}H_\perp/\Phi_0)^{1/2}}{2},
\end{equation}
where $n=1,2,\dots$ is the order of the Bragg resonance.

The dependencies $\frac{1}{2}k_\mathrm{VL}(H_\perp)$ for $n = 1$ and $k_\mathrm{VL}(H_\perp)$ for $n = 2$ are plotted in Fig. \ref{f2} d. As discussed, the positions of the bandgap center frequencies in the experiment coincide very well with the first two resonance conditions by Eq. (\ref{e2}) for Bragg scattering of spin waves from the periodic magnetic modulation induced by the vortex lattice. The shading in Fig. \ref{f2} d reflects the competing requirements posed on the magnon and fluxon subsystems for the observation of the bandgaps in the transmission spectra. Namely, at small values of $\mu_0H_\perp$, the vortex lattice is sparse and its crystallinity is poor. At the same time, small-$k_\mathrm{SW}$ SWs are easy to excite and they propagate over longer distances. In contrast, while the vortex lattice is characterized by a better crystallinity at larger $\mu_0H_\perp$, high-$k_\mathrm{SW}$ SWs are hard to excite and they are attenuated more rapidly. From this, it can be inferred that a clean observation of both bandgaps is only possible in the field range of $3 - 11$\,mT, where the vortex lattice is sufficiently uniform and spin waves can be excited efficiently.
\begin{figure}
\centering
    \includegraphics[width=0.93\linewidth]{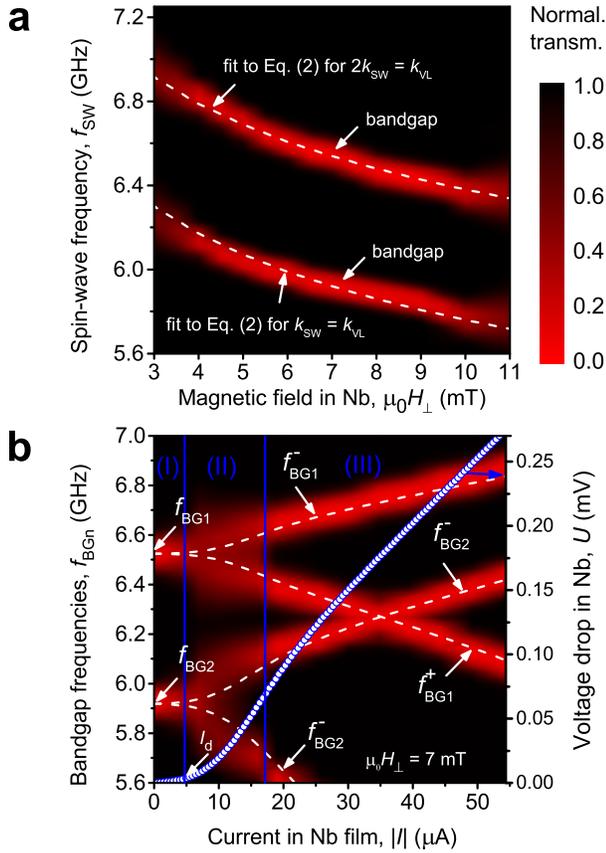}
    \caption{\textbf{Tailoring spin wave spectra by magnetic field and current}.
    \textbf{a} Normalized spin wave transmission through the Py/Nb bilayer as a function of the magnetic field $\mu_0 H_\perp$.
    The dashed lines are fits to Eq. (\ref{e2}) for the central bandgap frequencies.
    \textbf{b} Normalized spin wave transmission as a function of the absolute value of the current in the Nb layer.
    The positive current polarity corresponds to co-propagating spin waves and vortex lattice and vice versa.
    The dashed lines are fits $f^\pm_{\mathrm{BG}n}(k) = f(k_{\mathrm{BG}n} \pm \Delta k_n^{\pm})$ with $\Delta k_n^{\pm}= k (v_{SW}) - k (v_{SW}\pm v_{VL})$.
    In both panels $T = 8$\,K.
    \label{f3}
}
\end{figure}

To confirm this assumption, Fig.~\ref{f3} a shows a series of normalized transmission spectra through the Py waveguide as a function of $\mu_0 H_\perp$ in a contour plot together with dashed lines following Eq.~(\ref{e2}). Towards high fields ($\mu_0 H_\perp \gtrsim 10.5$\,mT) the second band gap becomes smeared and eventually vanishes. As can be seen from the green/red shading in Figs.~\ref{f2} b and d, this is explained by the decreasing excitation efficiency of spin waves by the antenna\cite{doi:10.1002/pssb.201147093}. Also for low fields, a smearing-out of the bandgaps is observed: As highlighted by the blue areas in Fig. \ref{f2} c and as discussed above, the long-range order in the vortex lattice is poor and the periodicity is not well defined at fields very close to the lower critical field. As can be seen from Fig.~\ref{f3} a, in the remaining field range, the position of the dips is described well by Eq. (\ref{e2}) and, hence, can be tuned readily by setting the field value. This shows that the fluxon-magnon interaction allows for the electrical characterization of the vortex lattice, and it also enables to create a reprogrammable magnonic crystal whose period can be adjusted by the vertical field component.


\textbf{Manipulation of spin waves by moving flux quanta.} The vortex lattice can be put into motion by an applied current to the Nb layer. The inelastic scattering of magnons with this moving lattice should allow for the observation of the spin-wave Doppler effect\cite{Vla08sci,Kra14pcm,Chu10prb}. The impact of a moving vortex lattice on the spin wave dynamics is summarized in Fig. \ref{f3} b. The color scale again shows the spin wave transmission amplitude, this time as a function of the spin wave frequency and the absolute value of the applied current. Please note that the plot includes a sweep of the currents in the positive (marked with $^+$) and negative ($^-$) direction. In addition, the blue-white circles show the current-voltage curve (CVC) of the Nb film. From the CVC, it can be extracted that below a certain critical current (depinning current) of about $I_\mathrm{d} = 5\,\mu$A, the voltage drop is essentially unchanged (region (I)). This is a typical characteristic property of a type II superconductor in the presence of a small transport current: The vortices are pinned, and only if the Lorentz Force $\mathbf{F}_\mathrm{L} = \mathbf{I} \times \mathbf{B}_\mathrm{VL}$ induced by the transport current $\mathbf{I}$ overcomes the pinning force $\mathbf{F}_\mathrm{p}$, the vortex lattice is set into motion \cite{Bra95rpp}. For lower currents, the vortices remain in their place and the spin wave spectrum is unaffected. In contrast, for larger currents, the moving vortices change the voltage drop and also the spin wave transmission characteristics. For $I\gtrsim I_\mathrm{d}$, the vortices are set into motion and for yet larger currents the flux flow regime is established, in which $v_\mathrm{VL}\propto|I|$. Accordingly, this part of the CVC can be split into two further regimes: For $I_\mathrm{d} < I < 17\,\mu$A, the vortex motion results in a change of the spin wave bandgaps and an overall blurring of the gaps (region (II)). In contrast, for $I > 17\,\mu$A, clean transmission spectra are recovered and the bandgap can be readily tuned by the applied current (region (III)).

These two different regimes can be associated with different regimes of vortex dynamics. In region (II), the vortex motion is not coherent. An unavoidable variation in the local pinning forces acting on individual vortices causes their depinning at slightly different current values. Consequently, the vortex lattice looses its long-range order until all vortices have been depinned\cite{Bra95rpp}. At larger currents (region (III)) which result in higher vortex velocities, the long-range order in the vortex lattice is recovered\cite{Kos94prl}. In this regime, the vortex lattice is characterized by a better crystallinity than at the depinning transition, and the bandgaps become once again well-defined.

For $I > 0$, the vortices move in the same direction as the spin waves propagate and $f_\mathrm{BG1}$ and $f_\mathrm{BG2}$ shift towards higher frequencies. In contrast, for $I < 0$,  the vortices and the spin waves travel in opposite directions and $f_\mathrm{BG1}$ and $f_\mathrm{BG2}$ shift towards lower frequencies. The magnitudes of the frequency shifts depend on the current polarity and they increase with increase of the current value. This behaviour can be understood on the basis of Bragg scattering accompanied by frequency shifts due to the Doppler effect. With the knowledge of the vortex velocity from the CVC and the knowledge of the spin-wave dispersion, it is possible to calculate the expected position of the Doppler-shifted bandgap. The dashed lines in Fig.~\ref{f3} b represent the calculated bandgap positions assuming coherent vortex motion. As is evident from the figure, in regime (III), the expected and the experimentally observed bandgap positions agree very well. Thus, the magnon spectrum can be used to detect and quantify the vortex motion as well as its coherency in such a magnon-fluxonic system. On the other hand, the fluxonic system allows for an externally tunable control of magnons by the application of ultra-low currents in the superconductor.
\enlargethispage{1\baselineskip}

In summary, we have demonstrated the fingerprints of inelastic magnon-fluxon interaction in a superconductor/ferromagnet heterostructure. The vortex lattice constitutes a tunable Bragg grating for spin waves, resulting in characteristic dips in the spin wave transmission. As the distance between vortices is inversely proportional to the square root of the magnetic field value, this allows for tuning the transmission bands and gaps in the magnon spectrum on demand. In turn, the magnon spectrum can be used to probe the presence and the periodicity of the Abrikosov lattice. In addition, we have demonstrated the magnon fluxon Doppler effect: A moving vortex lattice changes the position of the band gaps in the transmission spectrum. This results in a tunable control of magnons that also allows for a direct and straightforward electrical characterization of the vortex flow.

\section{Methods}

\subsection{Superconducting film growth and properties}
The Nb microstrip was fabricated by photolithography and Ar etching from an epitaxial (110) Nb film on a-cut sapphire substrate similar to the samples used in Ref. \cite{Dob17nsr}, but with a slower deposition rate resulting in better structural quality \cite{Dob12tsf}. Namely, the film was grown by dc magnetron sputtering in a setup with a base pressure in the $10^{-8}$\,mbar range. In the sputtering process the substrate temperature was 850$^\circ$C, the Ar pressure $4 \times 10^{-3}$\,mbar, and the growth rate was about\,0.5 nm/s. X-ray diffraction measurements revealed the (110) orientation of the film \cite{Dob12tsf}. The epitaxy of the film has been confirmed by reflection high-energy electron diffraction. The as-grown film has a smooth surface with an rms surface roughness of less than 0.2\,nm, as deduced from AFM scans in the range $1\,\mu$m$\times1\,\mu$m. The film is in the clean superconducting limit: Its room-temperature-to-10\,K residual resistance ratio is equal to about 50, the superconducting transition temperature is $8.82$\,K, and the upper critical field is estimated as $800$\,mT as deduced from fitting the dependence $\mu_0 H_\mathrm{c2}(T)$ to the phenomenological law $\mu_0 H_\mathrm{c2}(T)  =  \mu_0 H_{c2}(0)[1 - (T/T_\mathrm{c})^2]$. The background pinning in the as-grown Nb film is very weak and is characterized by a pinning activation energy of about $400$\,K as deduced from an Arrhenius analysis of the temperature dependences of the resistivity close to $T_\mathrm{c}$ in the limit of small dc current densities. After its growth, the Nb film was patterned into the four-probe geometry by photolithography in conjunction with Ar ion etching, forming a bridge with a length of $50\,\mu$m and a width of $4\,\mu$m. Two Au antennae, with an axis-to-axis distance of $5.5\,\mu$m, were fabricated on the sapphire substrate, spaced by $0.5\,\mu$m on both sides of the edges of the Nb bridge. The Nb layer was covered with a 5\,nm-thick SiO$_2$ cap layer.

\subsection{Ferromagnetic film growth and properties}
The Py waveguide was fabricated by means of a conventional lift-off technique and molecular beam evaporation. To prevent degradation of the magnetic properties on the surface, the Py waveguide was covered with a 5\,nm-thick Pt cap layer. Ferromagnetic resonance measurements (FMR) in a broad temperature range ($5$\,K to $300$\,K) were done on a reference Py $10\times10\,$mm$^2$ film with a microstrip antenna put on its surface. From the linewidth of the FMR resonance, a Gilbert damping parameter $\alpha_G$ of about $0.007$ was deduced. The effective magnetization value deduced is $M_\mathrm{eff}=676$\,kA/m. Calculations of the spin wave dispersion $k_\mathrm{SW}(f_\mathrm{SW})$ were done within the framework of the Kalinikos-Slavin theory adapted to a thin-film waveguide\cite{Kal86jpc,Bra17phr} for the dipole-exchange spin waves in ferromagnetic films. After growth, the Py film was patterned into a spin wave waveguide with a width of $2\,\mu$m by Ar ion beam etching.

\subsection{Measurements of microwave transmission}
The combined electrical voltage and microwave transmission measurements were made in a cryostat equipped with a superconducting solenoid using a vector network analyzer (VNA-FMR). The bias field angle was varied between $0^\circ$ and $10^\circ$ in the plane normal to the sample surface and parallel to the waveguide axis. The microwave signal of varying frequency from $300$\,kHz to $14$\,GHz was launched in the CPW by nonmagnetic SMP probe via semirigid, low-loss coaxial cables. The forward transmission coefficient (scattering parameter $S_{21}$, associated with the power received at port 2 relative to the power delivered to port 1) was measured by the VNA at the detector port. Two 100\,nm-thick and 500\,nm-wide Au antennae were produced on top of the waveguide $5.5\,\mu$m apart. The microwave stimulus was generated by the microwave source of the VNA, while the signal power was kept at $1\,\mu$W ($-30$\,dBm) being low enough to avoid nonlinear processes. Backward volume magnetostatic spin waves were detected by the second antenna while the excitation frequency was swept in the range from 6.5 to 10\,GHz. The depinning current was determined from current-voltage curves using a $10\,$nV voltage criterion.

\section*{Acknowledgements}
      The authors gratefully acknowledge financial support by the DFG in the framework of the Collaborative Research Center SFB/TRR-173 \textit{Spin+X} (Project B04).
           O.V.D. acknowledges the DFG for support through Grant No 374052683 (DO1511/3-1).
           O.V.D., V.V.K., R.V.V., V.A.S. acknowledge support from the European Commission within the framework of the program Marie Sklodowska-Curie Actions --- Research and Innovation Staff Exchange (MSCA-RISE) under Grant Agreement No. 644348 (MagIC).
					A.V.C. and T.B. acknowledge financial support within the ERC Starting Grant No. 678309 MagnonCircuits.
           Research leading to these results was also conducted within the framework of the COST Action CA16218 (NANOCOHYBRI) of the European Cooperation in Science and Technology.

\section*{Author contributions}
                                O.V.D., A.V.C., V.V.K., and V.A.S. conceived the experiment.
                                O.V.D. and A.V.C. designed the samples.
                                T.F. and R.S. fabricated the samples.
                                O.V.D. and R.V.V. performed the measurements.
                                O.V.D., T.B., and A.V.C. performed and evaluated the spin wave transmission simulations.
                                O.V.D. and A.V.C. led the project.
                                All authors discussed the results and co-wrote the manuscript.


%

\end{document}